\documentclass{myelsart} 
\pdfoutput=1
\usepackage{graphics}
\usepackage{epstopdf}
\usepackage{amssymb} 
\usepackage{amsmath}
\usepackage{cite}

\begin{document}

\begin{frontmatter}

\title{Loop-closure principles in protein folding}
\author{Thomas R.\ Weikl}
\address{Max Planck Institute of Colloids and Interfaces,  \\ Department of Theory and Bio-Systems, 14424 Potsdam, Germany}

\begin{abstract}
Simple theoretical concepts and models have been helpful to understand the folding rates and routes of single-domain proteins. As reviewed in this article, a physical principle that appears to underly these models is loop closure.
\end{abstract}


\end{frontmatter}

\section{Topology and loop closure} 

The topic of this review is the relation between the folding kinetics of proteins and their three-dimensional, native structures. Central questions concerning the folding kinetics are: How do proteins fold into their native structures, and what are the rates and routes of folding? Since their discovery in 1991 \cite{Jackson91}, two-state proteins have been in the focus of experimental studies \cite{Jackson98,Fersht99,Grantcharova01,Maxwell05}. These proteins fold from the denatured state to the native state without experimentally detectable intermediate states. The size of most two-state proteins is rather similar, roughly between 60 and 120 residues, with a few smaller or larger exceptions \cite{Jackson98,Grantcharova01,Kamagata04,Maxwell05}. Nonetheless, their folding rates range over six orders of magnitude: the fastest proteins fold on a microsecond \cite{Yang03,Kubelka04} and, if designed for speed, sub-microsecond time scale \cite{Xu06,Kubelka06}, whereas slow two-state proteins fold on a time scale of seconds \cite{Chiti99}. In 1998, Plaxco, Simons, and Baker \cite{Plaxco98} discovered that these folding rates correlate with a simple measure of the structural `topology', the relative contact order (CO). The relative CO is the average sequence separation $|i-j|$ of all contacts between amino acids $i$ and $j$ in the native structure, divided by the chain length. Proteins with many local contacts and, hence, small relative CO, tend to fold faster than proteins with many nonlocal, sequence-distant contacts and large relative CO. The discovery of Plaxco et al.\ pointed towards a `surprising simplicity' \cite{Baker00} in protein folding kinetics. The folding kinetics problem, i.e.~the problem of predicting folding rates and routes from native structures, appeared to be considerably simpler than the structure problem, the prediction of native structures from sequences, which requires detailed atomistic models \cite{Bradley05}.

The physical principle that underlies the correlation between folding rates and relative CO seems to be loop closure. Contacts with small CO can be formed by closing a small loop, which is fast and requires a small amount of loop-closure entropy, compared to closing a large loop \cite{Fersht00,Zhou04}. It seems plausible that protein structures with many local contacts form faster than proteins with more complex structures involving many nonlocal contacts, provided that the loop-closure entropies, or chain entropies, dominate over sequence-dependent interaction energies in the folding process. The strength of the correlation between folding rates and relative CO and related structural measures discussed in section 3 indicates such a dominance of topological or loop-closure aspects, at least for a majority of proteins. Depending on the considered set of two-state proteins, the absolute values $|r|$ of the Pearson correlation coefficients between folding rates and relative CO of two-state proteins vary between 0.75 and 0.9 (see Table 2).  The squares of these correlation coefficients range roughly from 0.6 to 0.8, which indicates that between 60 \% to 80 \% of the observed variations in the folding rates can be traced back to simple aspects of the overall structure or topology, rather than sequence-specific energetic aspects. 

Several experimental observations support the importance of protein topology and loop closure. First, insertion of small loops into turns of the protein structure slows down folding \cite{Ladurner97,Viguera97,Fersht00,Grantcharova00}. Second, inserting covalent crosslinks into the protein chain speeds up the folding process \cite{Otzen98,Grantcharova00,Schoenbrunner97,Camarero01,Ainavarapu07}. The crosslinks interconnect the chain and increase the localness of some of the contacts in the protein structure. Third, single-residue mutations that locally perturb energetic interactions typically have a `less than tenfold effect' \cite{Baker00} on the folding rate, which appears small compared to the variations in folding rate observed for two-state proteins.  For few single-residue mutants, larger changes in the folding rate have been observed \cite{Burton96,Yang03}.  Also, homologous proteins of the same size, which have the same structure but can differ considerably in sequence, have folding rates that differ typically by less than one or two orders of magnitude  \cite{Jackson98,Gunasekaran01,Zarrine05}, which appears, again, small compared to the six orders of magnitude observed for two-state proteins. 

\begin{figure}
\begin{center}
\resizebox{0.9\linewidth}{!}{\includegraphics{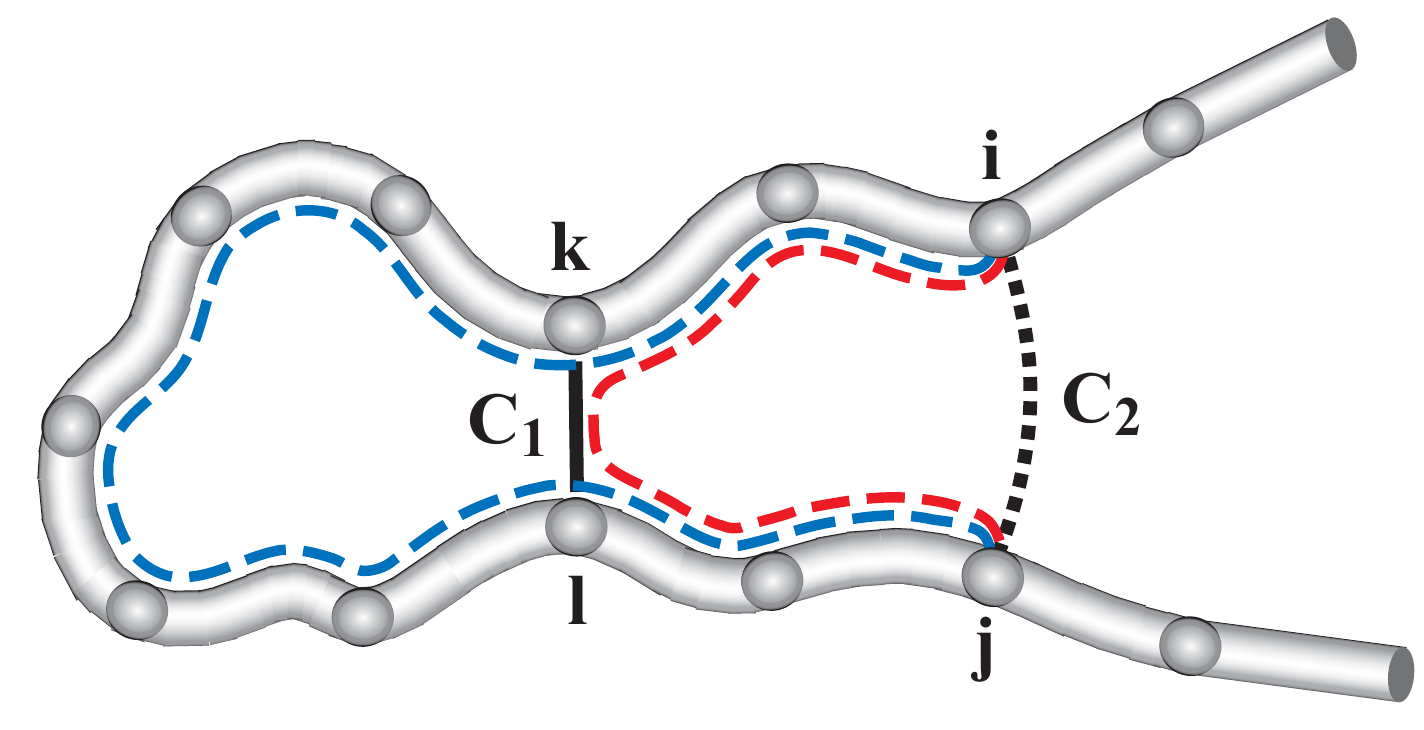}}
\end{center}
\vspace{1cm}
\caption{Loop lengths in partially folded conformations of a protein chain can be estimated via the graph-theoretical concept of effective contact order (ECO) \cite{Dill93,Fiebig93}. The ECO of the contact $C_2$ is the length of the shortest path between the two residues $i$ and $j$ forming the contact. The `steps' along this shortest path either are covalent bonds between adjacent residues, or noncovalent contacts formed previously in the folding process such as the contact $C_1$. In this example, the ECO for the contact $C_2$ is 5, since the shortest path (shown in red) involves two steps from $i$ to $k$, one step for the contact $C_1$ between $k$ and $l$, and two steps from $l$ to $j$. The contact order (CO), in contrast, is the sequence separation $|i-j|$ between the two residues, the number of residues along the blue path between $i$ and $j$. In this example, the CO of the contact $C_2$ is 10.  
\label{figure_ECO}}
\end{figure}

{ Can we predict folding routes from loop-closure principles?} The CO or sequence separation of a contact is the length of the loop that has to be closed to form the contact, provided that no other contacts have been formed prior that `short-circuit' the chain. In other words, the CO measures loop lengths for the fully unfolded state of the protein chain. But during folding, other contacts may have been formed prior to a specific contact between residues $i$ and $j$. The actual length of the loop that has be closed to form this contact in the partially folded state of the protein chain can be estimated via the graph-theoretical concept of effective contact order (ECO) \cite{Dill93,Fiebig93}. The ECO is the length of the shortest path between the two residues $i$ and $j$ that are brought in contact, see Fig.~\ref{figure_ECO}. The steps on this path either are bonds between neighboring residues in the chain, or contacts between residues that have been formed prior, such as contact $C_1$ between the residues $k$ and $l$ in Fig.~\ref{figure_ECO}. In contrast to COs, the ECOs thus are route-dependent: they depend on the sequence in which contacts are formed. On the minimum-ECO routes discussed in section 4, proteins fold, or `zip up', in sequences of events that involve only closures of small loops, which minimizes the entropic loop-closure barriers during folding. The minimum-ECO routes help to understand the shape of $\Phi$-value distributions from mutational analyses of the folding kinetics, see section 5.

\section{Contact maps, contact clusters, and topology}

To capture the concept of native-state topology more precisely, it is helpful to consider native contact maps. Contact maps are two-dimensional representations of three-dimensional protein structures. The native contact map of a protein is a matrix in which element $(i,j)$ indicates whether the residues $i$ and $j$ are in contact in the native structure. To some extent, the native contact map depends on the contact definition. In the map of Fig.~\ref{figure_contact_maps}(a), two residues are defined to be in contact if the distance between their backbone C$_\alpha$ atoms is smaller than the cutoff distance 7 \AA~chosen here, and in Fig.~\ref{figure_contact_maps}(b), if the distance between any of their non-hydrogen atoms is smaller than the cutoff distance 4.5 \AA. In the C$_\alpha$ contact map of Fig.~\ref{figure_contact_maps}(a), the contacts are arranged in clusters that correspond to the characteristic structural elements of CI2. These clusters are also present in the non-hydrogen-atom contact map of Fig.~\ref{figure_contact_maps}(b). In addition, the non-hydrogen-atom contact map contains more `isolated' contacts that mostly correspond to interactions of large sidechains, which are not represented in the backbone-centric C$_\alpha$ contact map. A third type of contact map is shown in Fig.~\ref{figure_contact_maps}(c). The different gray tones in this map indicate the numbers of contacting non-hydrogen atom pairs of two residues. This contact map is the basis for the calculation of the relative CO.

The contact maps of Fig.~\ref{figure_contact_maps} indicate the chain positions $i$ and $j$ of contacting amino acids, but not which of the twenty different types of amino acids are located at these positions. In other words, the contact maps do not contain sequence information, they just contain information on the structure. This structural information is rather detailed. Single-residue mutations can lead to deletion or addition of contacts, and homologous proteins of the same size can differ in many native contacts. Nonetheless, single-residue mutants and homologous proteins have the same overall structure. To capture the overall structure or `structural topology' of a protein, it is helpful to take a more coarse-grained view of contact maps and to focus on contact clusters, e.g.~on the clusters in the C$_\alpha$ contact map of Fig.~\ref{figure_contact_maps}(a). The size of contact clusters may vary between wildtype and mutants of a protein, or between homologous proteins of similar size. But the overall location of these clusters in the contact map in general stays the same. The contact clusters thus capture the overall structural topology of a protein.   

\section{Contact order, topological measures, and folding rates }

Simple measures of native-state topology are characteristic, average properties of contact maps. The relative CO defined by Plaxco et al.\ \cite{Plaxco98} is the average CO of all contacts between non-hydrogen atoms of the contact map shown in Fig.~\ref{figure_contact_maps}(c), divided by the chain length $N$. The CO of the contacting atoms simply is the sequence separation $|i-j|$ of the two residues $i\ne j$ in which the atoms are located. Depending on the data set, the obtained correlations between relative CO and the folding rates of two-state proteins vary between 0.75 and 0.92, see Table 1. Proteins with many local contacts between residues that are close in the chain sequence have a small relative CO and fold faster than proteins with many nonlocal contacts and large relative CO. The typically fast-folding $\alpha$-helical proteins have small relative COs since their contact maps contain many local, intra-helical contacts between residues $i$ and $i+3$ or $i+4$. Proteins with $\beta$-sheets, in contrast, have larger relative COs and, on average, fold slower. But also within the classes of $\alpha$-helical and $\beta$-sheet containing proteins, significant correlations between folding rates and relative CO can be observed \cite{Kuznetsov04}. 

Related topological measures that correlate with the folding rates of two-state proteins are the `long-range order' \cite{Gromiha01}, the `total contact distance' \cite{Zhou02}, and the number $Q_D$ of nonlocal contacts with CO $>$ 12 \cite{Makarov02,Makarov03} (see Table 1). The long-range order is the number of contacts with CO $>$ 12, divided by the chain length, and the total contact distance is the sum over the COs of all contacts, divided by the chain length squared. The topomer-search model of Makarov et al.\ \cite{Makarov02,Makarov03} predicts that the number $Q_D$ of nonlocal contacts is proportional to $\log k_f/Q_D$ where $k_f$ is the folding rate \cite{Makarov02,Makarov03}. The diffusive search for a topomer \cite{Debe99a,Debe99b,Makarov02,Makarov03}, i.e. for the ``set of unfolded conformations that share a common, global topology with the native state" \cite{Makarov03}, has been suggested as a physical principle that underlies the correlation between relative CO and folding rates \cite{Makarov02,Makarov03,Gillespie04}. { Recent extensive simulations with an off-lattice model indicate, however, that an unbiased diffusive search process for a native topomer ``would  take an impossibly long average time to complete" \cite{Wallin05}.}

Can topological measures capture the increase in folding rate that is caused by the insertion of covalent chain crosslinks \cite{Otzen98,Grantcharova00,Schoenbrunner97,Camarero01}?  
Inserting crosslinks such as disulfide bonds into the protein chain increases the localness of some of the native contacts, since the crosslinks `short-circuit' the chain. The natural extension of the CO of a contact in a crosslinked chain is  the {\em minimum} number of covalently connected residues between the two residues in contact. This minimum number is the ECO of the contact in the crosslinked but otherwise unfolded chain, and the relative ECO is the natural extension of the relative CO for a crosslinked chain. The relative ECO appears to overestimate changes in the folding rates caused by crosslinks \cite{Dixit06}. But a closely related pair of measures, the relative logCO and relative logECO, captures the folding rates of two-state proteins with and without crosslinks \cite{Dixit06}. The relative logCO is the average value for the logarithm of the CO of all contacts, divided by the logarithm of the chain length, and the relative logECO is the natural extension of this measure for crosslinked chains. 

Do topological measures also predict folding rates of non-two-state proteins? Non-two-state proteins exhibit at least one metastable, experimentally detectable intermediate state during folding. The correlations between relative CO and folding rates of protein sets that contain both two-state and non-two-state proteins are insignificant (see Table 2). For these protein sets, the correlation coefficients between absolute CO and folding rates are much larger than the correlation coefficients for the relative CO \cite{Ivankov03}, in contrast to two-state proteins (see Table 1 and 2). The absolute CO is the average CO of all native contacts, without the chain-length-dependent rescaling factor $1/N$ of the relative CO. Non-two-state proteins also exhibit strong correlations with the logarithm and the square root of the chain length $N$. 

There are also other simple models for protein folding rates. A conceptually somewhat different topological measure that correlates with protein folding rates is `cliquishness' \cite{Micheletti03}, which characterizes the overall clustering tendency of native contacts. In the past years, several groups have found that protein folding rates correlate with secondary structure content \cite{Gong03,Prabhu06,Huang07} or secondary propensity \cite{Ivankov04,Huang06}. It has recently been suggested that secondary structure determines protein topology \cite{Fleming06}. Native-state topology and loop closure thus may again be the principles that underlie the correlations between secondary structural measures and folding rates.  The topology and chain-length dependence of folding rates has also been studied in lattice \cite{Jewett02,Kaya03,Chan04,Faisca05a,Kachalo06} and simple off-lattice models of proteins \cite{Koga01,Cieplak03,Ejtehadi04,Wallin06}. { In lattice models, relatively strong correlations between folding rates and relative CO of model structures are observed if energetic terms that increase the folding cooperativity are included \cite{Jewett02,Kaya03,Chan04}.} 

\section{Folding routes and effective contact order}

The correlations between protein folding rates and simple topological measures inspired the development of statistical-mechanical models based on native-state topology. These models can be   grouped into three classes. First, there are models that use explicit representations of the protein chain with Go-type energy potentials \cite{Shea99,Clementi00,Clementi01,Clementi03,Hoang00,Koga01,Li01,Karanicolas02,Ejtehadi04,Brown04,Wallin06,Qin06}, named after the Japanese physicist Nobuhiro Go \cite{Taketomi75}. In these potentials, amino acids that are in contact in the native structure attract each other, while amino acids not in contact in the native structure repel each other, irrespective, at least to some extent, of the physical interactions between the amino acids. The second class of models assumes that amino acids can be in either of two states: native-like structured, or unstructured \cite{Alm99,Munoz99,Galzitskaya99,Garbuzynskiy04,Guerois00,Alm02,Bruscolini02,Karanicolas03,Henry04,Nelson06,Zwanzig95,Hilser96,Hilser06}. Partially folded states then are described by sets of structured amino acids. These models are inspired  by the Zimm-Bragg model for helix-coil transitions \cite{Zimm59}, which assumes that amino acids in helices can either be in a helix or a coil state. In the third class of models, partially folded states are characterized by the subset of native contacts formed in these states \cite{Weikl03a,Weikl03b,Weikl04,Weikl05,Shmygelska05}. \footnote{Other approaches that do not directly fall in one these three classes are the diffusion-collision model of Karplus and Weaver \cite{Karplus76,Karplus94,Islam02} as well as free-energy-functional \cite{Shoemaker97,Shoemaker99} and perturbed-Gaussian-chain methods \cite{Portman01,Kameda03}.}

\begin{figure}
\begin{center}
\resizebox{0.9\linewidth}{!}{\includegraphics{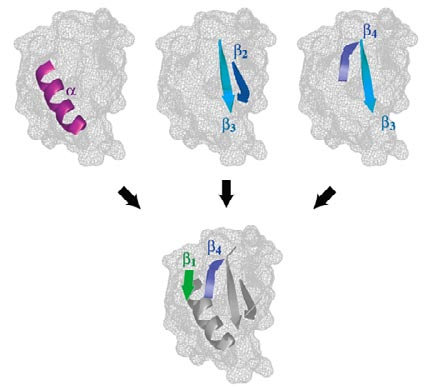}}
\end{center}
\vspace{1cm}
\caption{Minimum-ECO, or minimum-entropy-loss route of the protein CI2. Along this route, the strand pairing $\beta_1\beta_4$ is formed after the other three structural elements, the $\alpha$-helix and the strand pairings $\beta_2\beta_3$ and $\beta_3\beta_4$. The route minimizes the length of the loop that has to be closed to bring the two terminal strands $\beta_1$ and $\beta_4$ into contact (see Table 3).
\label{figure_CI2_route}}
\end{figure}

Folding routes can be predicted from native contact maps rather directly via the concept of effective contact order (ECO). The ECO is an estimate for the length of the loop that has to be closed to form a contact or contact cluster in a partially folded chain conformation (see Fig.~\ref{figure_ECO}). The contact clusters in a native contact map represent the characteristic structural elements of a protein. MD simulations indicate increased correlations between contacts of the same contact cluster \cite{Reich06}. In a coarse-grained view, individual folding routes can be described by the sequence in which contact clusters are formed. For the protein CI2, which has just four contact clusters, there are $4!=24$ possible sequences in which the clusters can be formed. The length of the loop that has to be closed to form a contact cluster in general depends on the sequence in which the clusters are formed. For example, the contact cluster $\beta_1\beta_4$ in the contact map of Fig.~\ref{figure_contact_maps}(a) represents contacts between the two terminal strands $\beta_1$ and $\beta_4$ of CI2. Forming this contact cluster from the fully unfolded state, i.e.~{\em prior} to the other three clusters, requires to close a relatively large loop of length 42 (see Table 3). However, forming $\beta_1\beta_4$ {\em after} the other three clusters $\alpha$, $\beta_2\beta_3$, and $\beta_3\beta_4$ requires only to close a relatively small loop of length 7. The reason is that the contacts of the clusters $\alpha$, $\beta_2\beta_3$, and $\beta_3\beta_4$  short-circuit the chain, which brings the two chain ends with the strands $\beta_1$ and $\beta_4$ into closer spatial proximity. On the minimum-ECO route \cite{Weikl03a,Weikl05}, i.e.~the folding route that minimizes the loop lengths and, thus, the entropic loop-closure barriers, the cluster  $\beta_1\beta_4$ is formed after the clusters $\alpha$, $\beta_2\beta_3$, and $\beta_3\beta_4$. On this route, $\alpha$, $\beta_2\beta_3$, and $\beta_3\beta_4$ form in parallel since the ECOs of these three clusters do not depend on the sequence in which they are formed (see Fig.~\ref{figure_CI2_route}). 

In general, there are two scenarios for two contact clusters (or structural elements) A and B of a protein. In the first scenario, the ECOs (or loop lengths) of the contact clusters A and B do not depend on the sequence in which the clusters are formed. The two clusters then are predicted to form {\em parallel} to each other. In the second scenario, the ECO of one of the two clusters, e.g. cluster B, is significantly smaller if cluster A is formed prior to B. The clusters are then predicted to form {\em sequentially}, provided that the total loop-closure cost for cluster B along this route, which includes the loop-closure cost for cluster A, is smaller than on other routes \cite{Weikl05}. An important point here is that the loop-closure dependencies between two contact clusters typically are strong in the second scenario, i.e.~the differences in loop lengths are large if the sequence of events in which the clusters are formed is reversed. Therefore, simple estimates of loop-closure entropies \cite{Weikl03a,Weikl04} or minimization of loop lengths \cite{Weikl05} are sufficient to derive the dominant minimum-ECO or minimum-entropy-loss routes. Detailed calculations show that the entropy loss for closing a loop in an unfolded chain is proportional to the logarithm of the loop length for large loops  \cite{Jacobson50,Chan90,Camacho95,Zhou01,Zhou04}, while the closure of short loops with up to about 5 residues is impeded by the chain stiffness \cite{Zhou01,Camacho95,Kellermayer97,Dietz06}. 
   
\section{$\Phi$-value distributions and topology}

Several experimental methods provide information on folding routes. The characterization of metastable, partially folded states of non-two-state proteins gives direct information on folding intermediates, provided these metastable states or `on-route' to the native state, and not `off-route' traps. Structural information on these intermediates can be obtained with hydrogen-exchange or NMR methods \cite{Englander00,Matthews93,Bai95,Juneja03,Krishna04,Bai06,Wales06}. 
Two-state proteins do not exhibit experimentally detectable, metastable intermediates during folding. Instead, the folding kinetics of many two-state proteins has been investigated via mutational analysis \cite{Itzhaki95,Villegas98,Chiti99,Ternstrom99,Kragelund99,Martinez99,Riddle99,Fulton99,Hamill00,Kim00,Mccallister00,Jaeger01,Otzen02,Northey02,Gianni03,Deechongkit04,Garcia04,Anil05,Wilson05,Petrovich06}. In a mutational analysis, a large number of mostly single-residue mutants of a protein is generated. For each mutant, the effect of the mutation on the folding dynamics is quantified by its $\Phi$-value \cite{Matouschek89,Fersht99}
\begin{equation}
\Phi= \frac{R T\ln( k_{\text{wt}}/k_{\text{mut}})}{\Delta G_{N}} \label{phi}
\end{equation}
Here, $k_{\text{wt}}$ is the folding rate for the wildtype protein, $k_{\text{mut}}$ is the folding rate for the mutant protein, and $\Delta G_{N}$ is the change of the protein stability induced by the mutation. The stability $G_{N}$ of a protein is the free energy difference between the denatured state $D$ and the native state $N$. In a more recent method termed $\Psi$-value analysis, divalent biHis metal-ion binding sites are introduced into a protein, and the folding rate of the mutant protein is studied as a function of the metal-ion concentration \cite{Krantz01,Sosnick04,Pandit06,Sosnick06}.

\begin{figure}
\begin{center}
\resizebox{0.65\linewidth}{!}{\includegraphics{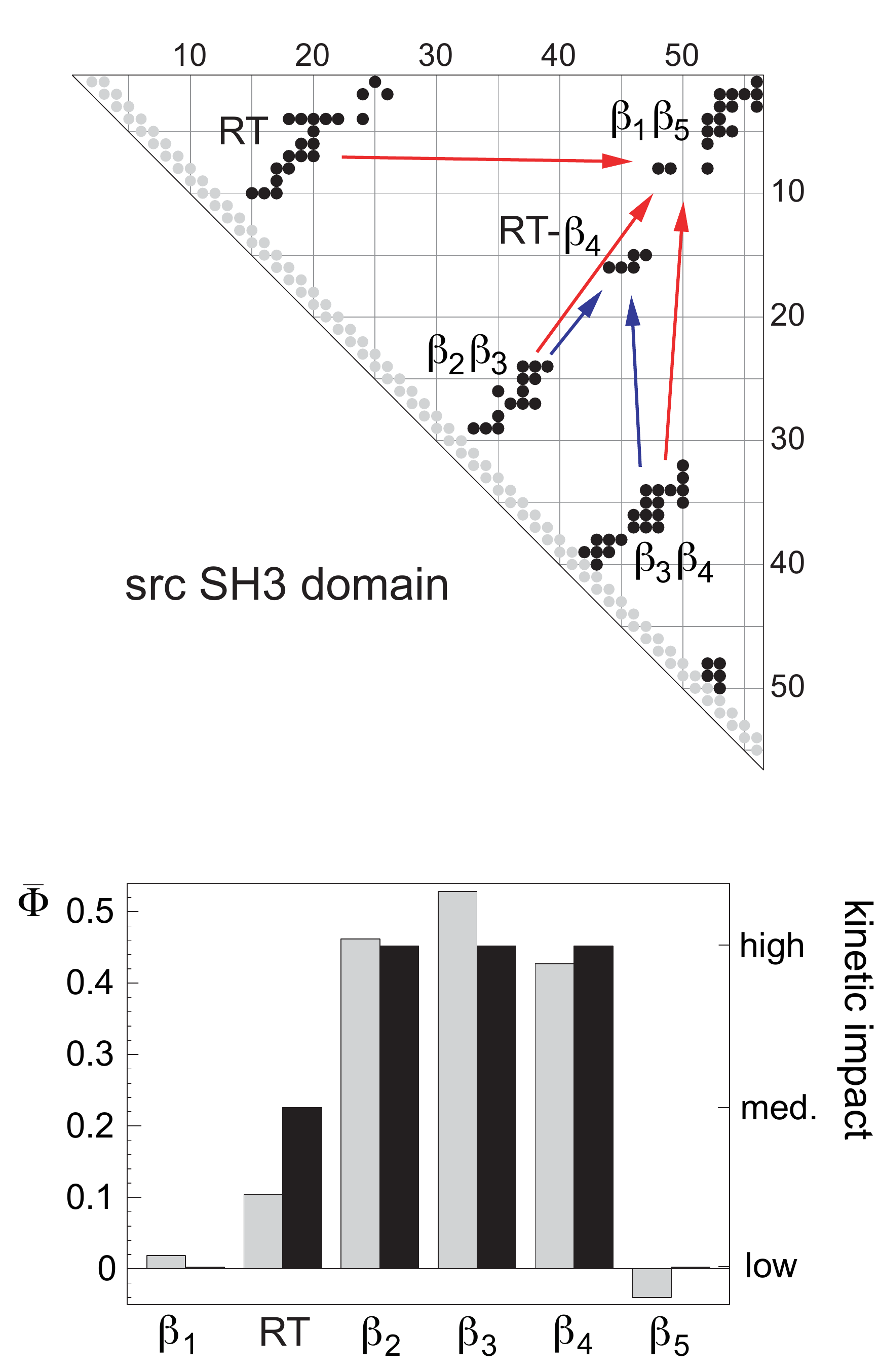}}
\end{center}
\vspace{1cm}
\caption{(Top) Minimum-ECO route of the src SH3 domain. The arrows indicate the sequences of events along this route. The red arrow pointing from the contact cluster RT to the cluster $\beta_1\beta_5$, for example, indicates that the RT loop is formed prior to the strand pairing $\beta_1\beta_5$. -- (Bottom) Average experimental $\Phi$-values \cite{Riddle99} for the $\beta$-strands and the RT loop (grey bars) and kinetic impact estimated from the minimum-ECO route (black bars). The kinetic impact of the strands $\beta_2$,  $\beta_3$, and $\beta_4$ is high since the clusters $\beta_2\beta_3$ and $\beta_3\beta_4$ are formed prior to both nonlocal clusters RT-$\beta_4$ and  $\beta_1\beta_5$ \cite{Weikl05}. The kinetic impact of RT is medium since the cluster is formed prior to only one of the nonlocal clusters,  $\beta_1\beta_5$ . The kinetic impact of $\beta_1$ and $\beta_5$ is low since the cluster $\beta_1\beta_5$ forms last, parallel to RT-$\beta_4$.
\label{figure_srcSH3}}
\end{figure}

$\Phi$-values have been calculated in statistical-mechanical models that are based on native structures
 \cite{Alm99,Alm02,Munoz99,Galzitskaya99,Clementi00,Guerois00,Li01,Kameda03,Karanicolas02,Clementi03,Brown04,Garbuzynskiy04} and from Molecular Dynamics unfolding simulations at elevated temperatures \cite{Li96,Lazaridis97,Day05}.  The detailed modeling of $\Phi$-values requires estimates for mutation-induced free energy changes \cite{Merlo05,Weikl07}, which goes beyond simple topology-based modeling. However, on a more coarse-grained level, the level of average $\Phi$-values for secondary structural elements, important aspects of $\Phi$-value distributions are captured by native-state topology. An average $\Phi$-value close to zero for a secondary structural element (i.e.~a helix or a $\beta$-strand) indicates that mutations in the secondary element affect the folding rate only marginally, see eq.~(\ref{phi}). In contrast, a large average $\Phi$-value indicates that mutations have a strong impact on the folding rate. In a sense, the average $\Phi$-values thus capture the `kinetic impact' of secondary elements. The kinetic impact can also be estimated from minimum-ECO routes. The minimum-ECO route of the src SH3 domain is shown in Fig.~\ref{figure_srcSH3}. Here, an arrow pointing from a contact cluster A to a cluster B in the contact map indicates that A is formed prior to B.  On the minimum-ECO route, the contact cluster RT-$\beta_4$ forms after $\beta_2\beta_3$ and $\beta_3\beta_4$, and the cluster $\beta_1\beta_5$ after RT, $\beta_2\beta_3$ and $\beta_3\beta_4$. The clusters RT-$\beta_4$ and $\beta_1\beta_5$ are nonlocal clusters and form in parallel on this route. The other three large contact clusters, the RT loop, an irregular, hairpin-like structure, and the two $\beta$-hairpins $\beta_2\beta_3$ and $\beta_3\beta_4$, are local clusters. Local clusters contain contacts with small CO and, thus, are located close to the diagonal of the contact map. 

The kinetic impact of a contact cluster can be estimated from how often the cluster appears along the minimum-ECO route to other clusters \cite{Weikl05}. The kinetic impact here is a semi-quantitative concept and can attain the values high, medium, or low. The kinetic impact of $\beta_2\beta_3$ and $\beta_3\beta_4$, for example, is high since these clusters appear on the route to both nonlocal clusters. The kinetic impact of RT is medium since the cluster only appears on the route to $\beta_1\beta_5$. The kinetic impact of  $\beta_1\beta_5$ and, thus, of the strands $\beta_1$ and $\beta_5$ is low since this cluster form last. The kinetic impact  derived from the minimum-ECO route agrees with average $\Phi$-values for the secondary elements (see Fig.~\ref{figure_srcSH3}). The $\Phi$-value distribution of the src SH3 domain is {\em polarized}, i.e.~the average $\Phi$-values are large for some of the secondary elements (the strands $\beta_2$,  $\beta_3$, and $\beta_4$), and small for others (the strands $\beta_1$ and $\beta_5$). The agreement between average $\Phi$-values and estimated kinetic impact shows that the polarized shape of the  $\Phi$-value distribution can be understood from simple topology-based modeling. 

The $\Phi$-value distributions of two-state proteins are either polarized or diffuse. In a {\em diffuse} distribution, the average $\Phi$-values for the secondary elements are of similar magnitude and differ not more than, say, a factor 2 from each other.   A diffuse distribution of kinetic impact occurs, e.g., if all clusters are involved on an `equal footing' in the formation of a single rate-limiting cluster. On the minimum-ECO route of CI2, for example, $\beta_1\beta_4$ forms after the other three clusters, which results in a diffuse distribution of kinetic impact, in agreement with the experimental $\Phi$-value distribution \cite{Weikl05}. A polarized distribution, in contrast, occurs if some clusters have a central role on the minimum-ECO route, such as the clusters $\beta_2\beta_3$ and $\beta_3\beta_4$ of the src SH3 domain. Also for other two-state proteins, the diffuse of polarized shape of their $\Phi$-value distributions can be traced back to minimum-ECO routes and native-state topology \cite{Weikl05}.

Minimum-ECO routes also help to understand why `topological mutations' such as circular permutations can have a drastic impact on the $\Phi$-value distribution \cite{Weikl03a}. In a circular permutation, the chain ends of a protein are covalently connected, and the chain is `opened' at a different location \cite{Viguera96,Otzen98,Lindberg01,Lindberg02,Miller02,Bulaj04,Lindberg06}. The protein still folds into the same structure \cite{Viguera96}, but the `rewiring' of the protein chain changes the loop-closure connections between the structural elements, and hence the minimum-ECO routes. The effect of circular permutations on folding routes and $\Phi$-value distributions has also been studied with protein models that use explicit chain representations and Go-type potentials \cite{Clementi01,Hubner06}.

The prediction of minimum-ECO routes is purely topology-based, i.e.~it is based on native contact clusters and the entropic loop-closure dependencies between these clusters. Sequence-dependent, energetic effects may play a role if parallel processes on minimum-ECO routes have a similar loop-closure cost. The sequence-dependent energies then can lead to a situation where one of these processes dominates the folding kinetics \cite{Weikl05}. 

\section{Summary and outlook}

The kinetic protein folding principle considered in this article is loop closure. Evidence for such a principle comes from strong correlations between protein folding rates and topological measures, and from experiments in which the effect of loop insertions \cite{Ladurner97,Viguera97,Fersht00,Grantcharova00} or crosslinks \cite{Otzen98,Grantcharova00,Schoenbrunner97,Camarero01,Ainavarapu07} on the kinetics is studied. Simple `zipping' or minimum-ECO models that are based on the loop-closure relations between structural elements help to understand the shape of $\Phi$-value distributions from mutational analysis of the folding kinetics \cite{Weikl05}. 

We have focused here an kinetic models and measures that are based on native-state topology. In the past years, the microsecond folding of ultrafast proteins has also been studied in Molecular Dynamics simulations with physical, atomistic force fields \cite{Duan98,Snow02,Snow04,Settanni05}.  An ultimate goal is to model both native structures and folding kinetics with such physical, sequence-based force fields. Even if this goal is achieved, the question remains whether there are simple principles that govern protein folding, such as the loop-closure principle considered here. In the next decade(s), detailed atomistic folding trajectories of two-state proteins may help to assess and establish such principles.

\subsection*{Acknowledgements}

I would like to thank Ken Dill for inspiring discussions, and Purushottam Dixit, Matteo Palassini, and Lothar Reich for fruitful collaborations.

\clearpage

\begin{table}
Table 1:  Correlation coefficients $|r|$ between folding rates of two-state proteins 
  and simple topological measures

\begin{center}
\hspace*{-1.5cm}
\begin{tabular}{c|c|c|cccccc}
 & & size of & \\[-0.15cm]
Authors & Ref. &  protein set & relCO & absCO  & LRO & TCD & $Q_D$ & logCO \\ 
\hline
Plaxco et al. & \cite{Plaxco00}, \cite{Makarov03} &  24 & 0.92 & & & & 0.88 & \\
Gromiha \& Selvaraj & \cite{Gromiha01} & 23 & 0.79$^{(a)}$ &  &  0.78 \\
Zhou \& Zhou & \cite{Zhou02} & 28 & (0.74)$^{(b)}$ & &  0.81 & 0.88 \\ 
Micheletti & \cite{Micheletti03} & 29 & 0.75 & 0.70 \\
Ivankov et al.\ & \cite{Ivankov03} & 30 & 0.75 & 0.51 \\
Kamagata et al.\ & \cite{Kamagata04} & 18 & 0.84 & 0.78 & & & 0.88 \\
Dixit \& Weikl & \cite{Dixit06} & 26 & 0.92 & 0.69 & 0.84$^{(c)}$ & 0.90$^{(c)}$ & 0.82$^{(c)}$ & 0.90 \\
\end{tabular}
\end{center}
~\\[0.5cm]
Absolute values $|r|$ of the Pearson coefficient for the correlations between folding rates of several sets of two-state proteins and relative contact order (relCO) \cite{Plaxco98}, absolute contact order (absCO) \cite{Grantcharova01}, long-range order  (LRO) \cite{Gromiha01}, total contact distance (TCR) \cite{Zhou02}, and logCO \cite{Dixit06}. In case of the number of nonlocal contacts $Q_D$ \cite{Makarov02}, the given coefficients report the correlations between $Q_D$ and $\log k_f/Q_D$ where $k_f$ are the folding rates.

----------- \\
$^{(a)}$ calculated from table 1 of Ref.~\cite{Gromiha01} \\
$^{(b)}$ the value is given in brackets since a slightly different definition for the relative CO has been used \\
$^{(c)}$ the values have been calculated for the protein structures given in Ref.~\cite{Dixit06}. The protein set is the set of Grantcharova et al.\ \cite{Grantcharova01}, which extends the set of Plaxco et al.\ \cite{Plaxco00} by two proteins.
\end{table}

\clearpage

\begin{table}
Table 2:  Correlation coefficients $|r|$ for folding rates of sets of two-state and non-two-state proteins

\begin{center}
\begin{tabular}{c|c|c|cccc}
 && size of & \\[-0.15cm]
Authors & Ref. &  protein set & relCO & absCO  &  $\ln(N)$ & $N^{1/2}$ \\
\hline
Ivankov et al., Li et al. &  \cite{Ivankov03}, \cite{Li04} & 57 & 0.1&  0.74 & 0.72  & 0.71 \\
Naganathan \& Mu\~noz & \cite{Naganathan05} & 69 & &  & &  0.74   \\ 
Kamagata et al.\ & \cite{Kamagata04}  & 40 & 0.09 & 0.72 & 0.68 & 0.67
\end{tabular}
\end{center}
~\\[0.5cm]
Absolute values $|r|$ of the Pearson coefficient for the correlations between folding rates of sets of two-state and non-two-state proteins and relative contact order (relCO), absolute contact order (absCO), and the logarithm and square root of the chain length $N$. The set of Naganathan and Mun\~oz  \cite{Naganathan05} extends the set of Ivankov et al.\ \cite{Ivankov03} by 12 proteins. The correlation coefficients for the set of Kamagata et al.\  have been calculated from the data given in Table 1 and 2 of Ref.~\cite{Kamagata04}.
\end{table}


\begin{table}
Table 3:  Loop lengths (ECOs) for forming the strand pairing $\beta_1\beta_4$ of CI2

~
\begin{center}
\begin{tabular}{c|c}
structural elements  & minimum ECO \\[-0.15cm]
formed prior  & for $\beta_1\beta_4$ \\
\hline
$\alpha + \beta_2\beta_3 + \beta_3\beta_4$  & 7 \\ 
$\alpha + \beta_2\beta_3$ & 12 \\
$\beta_2\beta_3 + \beta_3\beta_4$ & 19 \\
$\alpha +  \beta_3\beta_4$  & 24 \\ 
$\beta_2\beta_3$ & 23 \\
$\alpha$ & 31 \\
$\beta_3\beta_4$ & 36 \\
-- & 42 
 \end{tabular}
\end{center}

~\\
The given ECOs are the minimum ECOs among all contacts of the cluster $\beta_1\beta_4$. The contact clusters are defined as in Ref.~\cite{Weikl05}.
\end{table}

\clearpage

%


\begin{figure}
\begin{center}
\resizebox{0.6\linewidth}{!}{\includegraphics{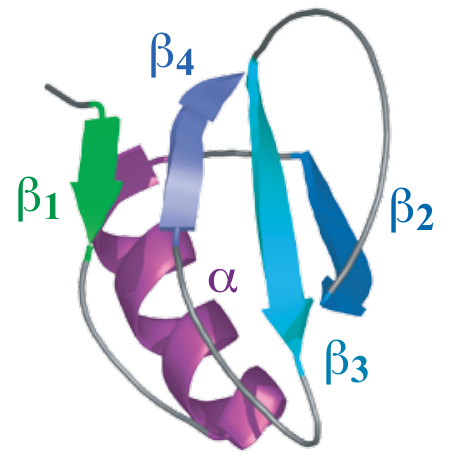}}
\end{center}
\vspace{1cm}
\caption{The structure of the protein CI2 consists of an $\alpha$-helix packed against a four-stranded $\beta$-sheet \cite{McPhalen87}.}
\label{figure_CI2_structure}
\end{figure}
\clearpage

\begin{figure}
\vspace{-1cm}
\begin{minipage}{0.55\linewidth}
\resizebox{\linewidth}{!}{\includegraphics{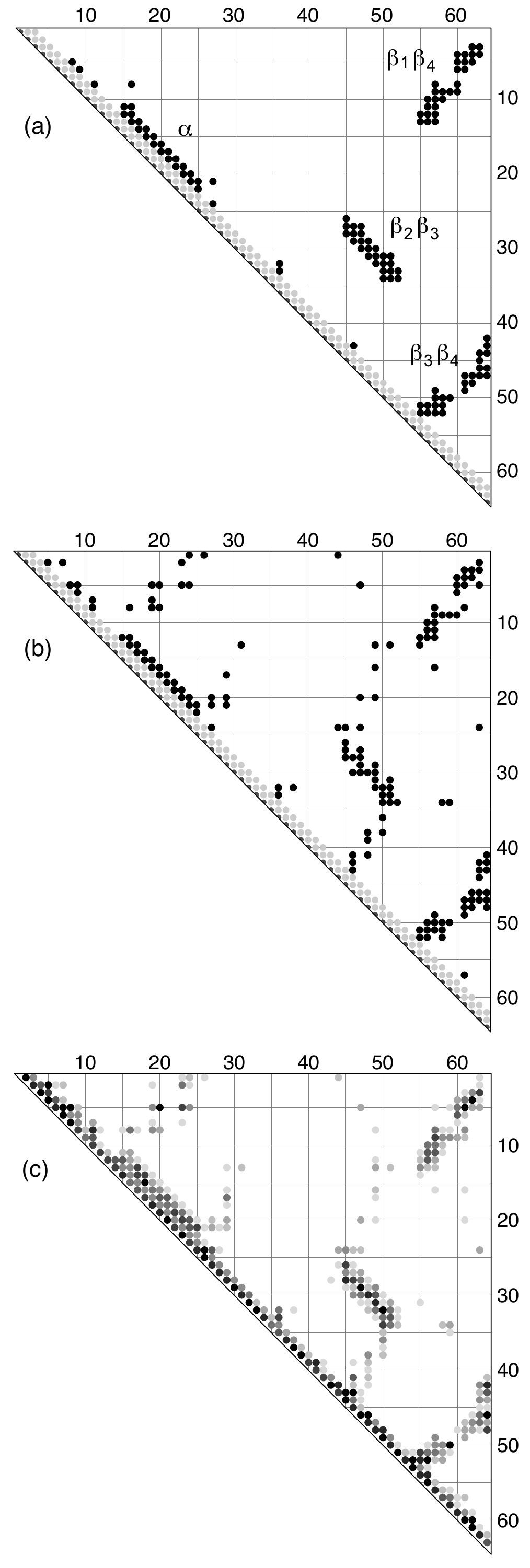}}
\end{minipage}
\hspace*{1cm}
\begin{minipage}{0.45\linewidth}
\caption{Native contact maps of the protein CI2 shown in Fig.~\ref{figure_CI2_structure}, for different contact definitions: (a) A black dot at position $(i,j)$ indicates that the C$_\alpha$ atoms of the residues $i$ and $j$ are within the cutoff distance 7 \AA. The residue numbers are specified along the two axes of the contact map. The four large clusters of contacts represent the structural elements of CI2, i.e.~the $\alpha$-helix and the three $\beta$-strand pairings $\beta_2\beta_3$, $\beta_3\beta_4$, and $\beta_1\beta_4$. -- (b) Black dots $(i,j)$ indicate that at least two non-hydrogen atoms of the residues $i$ and $j$ are within cutoff distance 4.5 \AA. As above, contacts $(i,j)$ of neighboring or next-nearest neighboring residues with $|i-j|\le 2$ (gray dots along the diagonal) are not taken into account. -- (c) The gray scale of the dots indicates the numbers of non-hydrogen-atom pairs of two residues $i$ and $j$ that are within the cutoff distance 6 \AA. Black dots represent residue pairs for which more than 40 different non-hydrogen-atom pairs are in contact, lighter gray colors represent residue pairs with fewer non-hydrogen-atom contacts.
\label{figure_contact_maps}}
\end{minipage}
\end{figure}
%


%


\end{document}